\documentclass[aps,prl,twocolumn,showpacs,preprintnumbers]{revtex4}
\usepackage{amsfonts}

\usepackage[pdftex]{graphicx}
\usepackage{dcolumn}
\usepackage{bm}
\include{biblio}

\begin{document}

\title{High-power Soliton-induced Supercontinuum Generation and Tunable
Sub-10-fs VUV Pulses from Kagome-lattice HC-PCFs}
\author{Song-Jin Im$^{1,2}$}
\author{Anton Husakou$^{1}$}
\author{Joachim Herrmann$^{1}$}
\affiliation{$^1$Max-Born-Institute for Nonlinear Optics and Short Pulse Spectroscopy,
Max-Born-Str. 2a, D-12489 Berlin, Germany}
\affiliation{$^2$Natural Science Center, KimIlSung University, Daesong District, Pyongyang, DPR Korea}
\begin{abstract}
We theoretically study a novel approach for soliton-induced high-power
supercontinuum generation by using kagome lattice HC-PCFs filled with a
noble gas. Anomalous dispersion and broad-band low loss of these fibers
enable the generation of two-octave broad spectra by fs pulses, with high
coherence and high spectral peak power densities up to five orders of
magnitude larger than in standard PCFs. In addition, up to 20\% of the
output radiation energy forms a narrow UV/VUV band, which can be tuned by
controlling the pressure in the range from 350 nm to 120 nm. In the temporal domain this corresponds to 
sub-10-fs UV/VUV pulses with pulse energy of few tens of $\mu$J, 
caused by the formation of a high-order soliton emitting non-solitonic radiation.
\end{abstract}
\maketitle

The discovery of photonic crystal fibers (PCF) had a strong impact on the
field of nonlinear fiber optics and led to numerous physical and
technological advances and interdisciplinary
applications \cite{russell_science_2003}. One of the most interesting type of PCFs is the
hollow-core PCF (HC-PCF) \cite{cregan_science_1999} in which light is guided
in the hollow core via the bandgap in the periodical cladding consisting of
a two-dimensional array of air holes. These fibers offer low loss
coefficients enabling diverse applications in nonlinear optics. However, the main drawback of these
fibers is their intrinsically narrow transmission bandwidth determined by
the bandgaps.

\begin{figure}
\includegraphics{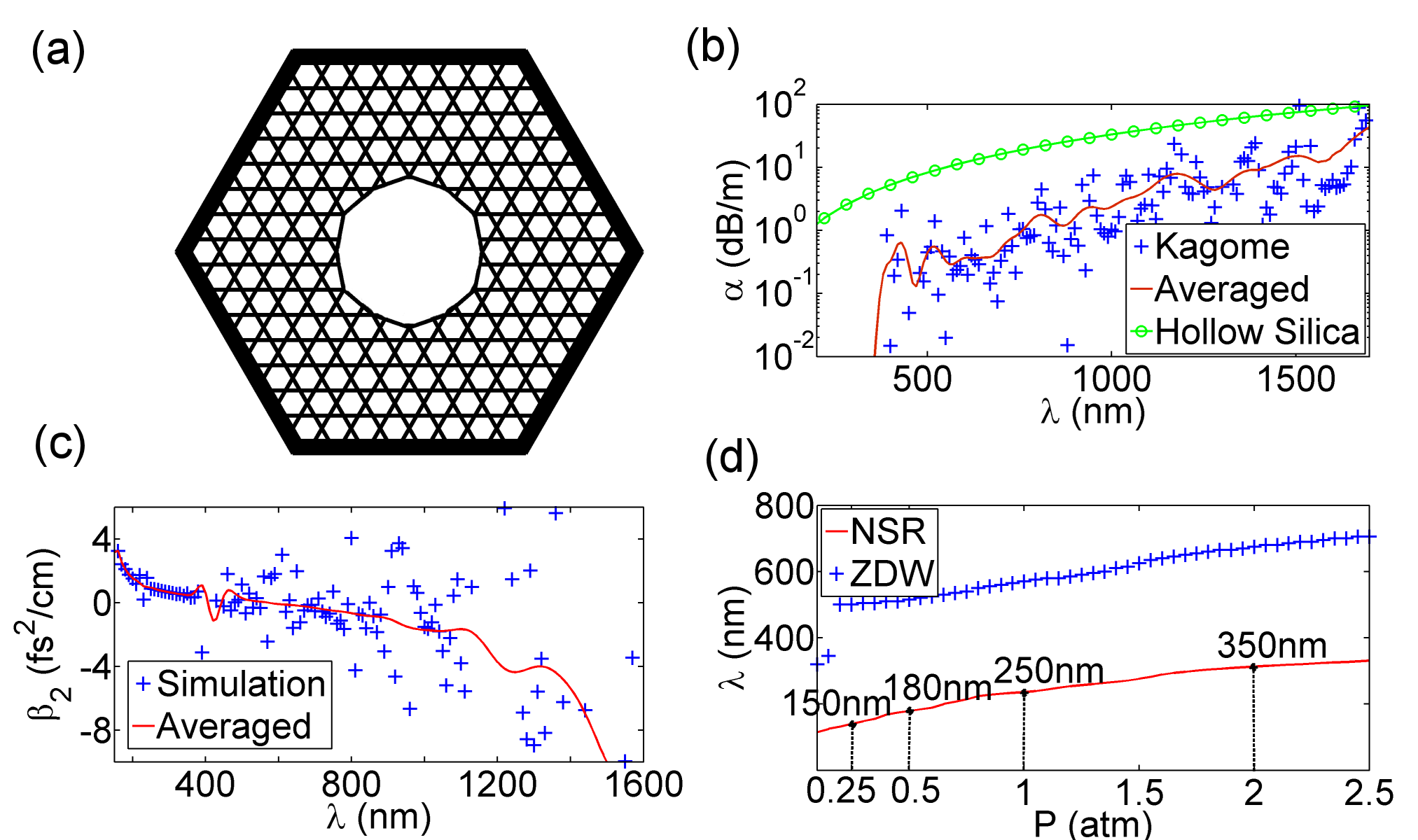}
\caption{Cross-section (a), loss (b), GVD (c) and zero-dispersion wavelength
(ZDW) and phase-matching wavelength (d) of a 3-ring-core kagome lattice HC-PCF. In (b)
and (c), a 3-ring-core kagome lattice HC-PCF with a lattice pitch of 16 $%
\protect\mu$m and a strut thickness of 0.2 $\protect\mu$m filled with argon
at 1 atm is considered. In (b), the blue crosses represent the direct
numerical simulations, the red solid curve is the result after averaging over 
inhomogenities and the green circles are the loss of a hollow silica
waveguide with the same core diameter. In (c), the blue crosses represent
the direct numerical simulations and the red solid curve is the averaged results.
In (d), the dependence of ZDW (blue crosses) and the phase-matching
wavelength (red solid curve) on the gas pressure are presented.}
\label{fig:1_Kagome}
\end{figure}

An alternative HC-PCF design replaces the triangular lattice of circular air
holes with a kagome lattice [see Fig. \ref{fig:1_Kagome}(a)] \cite%
{benabid_science_2002,couny_optlet_2006,couny_science_2007}. The photonic guidance of this fiber is not based on a bandgap but on the inhibited coupling between the core and cladding modes. Due to this guiding mechanism kagome lattice
HC-PCFs exhibit broadband transmission with a loss lower than 1 dB/m
covering the spectral range from the infrared up to the VUV. 
These fibers exhibit controlled anomalous dispersion for
UV or visible wavelengths both for 1-cell-core as well for 3-ring-core
geometries for core diameters in the range of 10 $\mu$m to 80 $\mu$m \cite%
{im_optexp_2009}.

As will be shown in this letter, the combination of broadband transmission
and control of zero-dispersion wavelength from visible to UV is of great
interest for applications in ultrafast nonlinear optics, especially for a
novel type of high-power coherent white light sources (supercontinuum, SC).
It is well known that lasers generate high-brightness coherent light only
within a restricted bandwidth, while many applications require much broader
coherent light with high brightness. A breakthrough in the development of
coherent white-light sources was the discovery of two-octave-spanning
supercontinuum generation in solid-core PCFs by using femtosecond pulses
with only nJ energy from a modelocked laser \cite%
{ranka_optlet_2000}.
The dramatic spectral broadening of low-energy femtosecond pulses in
solid-core PCFs significantly differs from previously known spectral
broadening mechanisms; it is related to the soliton dynamics in the
anomalous-dispersion region of the PCFs \cite{husakou_PRL_2001}. Subsequent
experimental studies by many groups provided evidence of this
soliton-induced mechanism of SC generation \cite%
{herrmann_PRL_2002,dudley_optexp_2002,ortigosa-blanch_joptsocam_2002}, for a
review see \cite{dudley_revmodphys_2006}. Supercontinuum generation from
solid-core PCFs plays a key role in a wide range of applications, such as in
optical frequency metrology, absorption spectroscopy, optical coherence
tomography and biophysics.

Despite the great progress in this field, a significant challenge is to
increase the available SC peak power and to extend the spectral broadening
into the ultraviolet and vacuum ultraviolet region; 
advances in this direction are requested by many applications.
Unfortunately, the small radii in solid-core PCFs and material damage
severely limit the SC peak power densities to tens of W/nm in these fibers.
In this paper we propose and theoretically study a novel approach for
high-power optical SC generation in argon-filled kagome lattice HC-PCFs. We
predict that in such waveguides high-coherence two-octave broad spectra, with
up to five orders of magnitude higher spectral peak power densities than in
solid-core PCFs, can be generated. It is enabled by three advantages of
kagome-lattice argon-filled HC-PCFs: dispersion control by pressure and
anomalous dispersion in the visible and UV, large core diameters in the
range from 10 $\mu$m to 80 $\mu$m, and high ionisation threshold of argon.
The underlying mechanism differs in an important aspect from that in solid-core
PCFs where the supercontinuum arises by the emission of several fundamental
solitons. In contrast, in the case of a kagome lattice HC-PCF it arises from a single high-order soliton. 
The reason for this phenomenon is the absent Raman effect in
argon and the low third-order dispersion which leads to a stable propagation
of a higher-order soliton over much longer propagation lengths.

A second predicted phenomenon is that the output radiation contains a
sub-10-fs UV/VUV pulse which carries about 20\% of input energy corresponding to
few tens of $\mu$J. The spectrum of this pulse is a narrow-band UV/VUV peak
which can be tuned by pressure variation in the range of 350-120 nm. 
These high-energy VUV pulses can be identified
as the non-solitonic radiation from a high-order soliton at the stage of
maximum compression that possesses the highest amplitude and the broadest
bandwidth. Note that up to now no standard method for the generation of ultrashort
pulses in the VUV range exist, and only relatively modest results compared
with the progress in the near-infrared spectral range has been achieved,
remarkable here is e.g. the generation of 11-fs pulses at 162 nm with 4 nJ energy 
\cite{kosma_optlet_2008} or 160-fs pulses with a higher energy of 600 nJ at
161 nm \cite{tzankov_optexp_2007}. However, ultrashort pulse sources in the
UV and VUV spectral range are essential tools in many applications, in
particular in time-resolved spectroscopy of molecules, requiring further
progress and alternative methods in this field.

Recently, two of the authors studied dielectric-coated metallic hollow
waveguides as an option for the generation of high-power high-coherent
supercontinua \cite{husakou_optexp_2009}. However, the challenges in the
fabrication of these fibers require to proceed with the search for a viable
type of hollow waveguides enabling dispersion control and high-power SC
generation.

The cross-section of the studied kagome lattice HC-PCF with a 3-ring-core is
presented in Fig. \ref{fig:1_Kagome}(a), in which a hollow core filled with a
noble gas is surrounded by a kagome lattice cladding and a bulk fused silica
outer region. For the calculation of the propagation constant $\beta (\omega
)$ and loss $\alpha (\omega )$ of this waveguide the finite-element Maxwell
solver JCMwave was utilized \cite{im_optexp_2009}. The dispersion of fused
silica as well as that of argon was described by the Sellmeyer formula for
the corresponding dielectric function. In Fig. \ref{fig:1_Kagome}(b) the loss coefficient is
presented which is few orders of magnitude lower than the one of a hollow
silica waveguide with the same core diameter, it decreases with decreasing
wavelengths and has a magnitude of about 1 dB/m at 800 nm. This low loss is
mainly influenced by the strut thickness in the kagome lattice cladding and
only weakly depend on the core diameter. We have found that the guiding
range of the kagome-lattice HC-PCF in the UV/VUV range is limited only by
the loss of argon, while the intrinsic silica loss in the 120-200 nm range does not
lead to high waveguide loss.

Figure \ref{fig:1_Kagome}(c) demonstrates the possibility to achieve
anomalous GVD at a desired wavelength by choosing a radius smaller than 100 $%
\mu$m. In a real waveguide, there are longitudinal variations of the
structure parameters due to manufacturing imperfections, leading to fast
longitudinal variation of the propagation constant, which however will be
smoothed out during propagation. This smoothing can be also performed in the
frequency domain, since the position of the spikes in the loss and
dispersion curves scales correspondingly with the varying structure parameters.
We assume a 5\%
variation depth of the inhomogeneity and consider an averaged loss and GVD,
as depicted by the red solid curves in Fig. \ref{fig:1_Kagome}(b),(c). 
In Fig. \ref{fig:1_Kagome}(d) by the blue crosses the dependence of
the zero-dispersion wavelength (ZDW) on the gas pressure is shown. One can
see that the ZDW can be tuned from 700 to below 500 nm by varying the
pressure from 2.5 atm to 0.25 atm. 

For the numerical simulations we use a generalized version of the
propagation equation for the electric field strength $E$ of forward-going
waves \cite{husakou_PRL_2001,husakou_optexp_2009}
\begin{eqnarray}
\frac{\partial E(z,\omega)}{\partial z}=i\left(\beta(\omega)-\frac{\omega}{c}%
\right)E(z, \omega)-\frac{\alpha(\omega)}{2}E(z,\omega)+  \nonumber \\
\frac{i\omega^2}{2c^2\epsilon_0\beta_{j}(\omega)}P_{NL}(z,\omega)
\label{eq:one}
\end{eqnarray}
where $P_{NL}$ describes the nonlinear Kerr polarization as well as the
photoionization-induced nonlinear absorption and phase modulation by the plasma,
and $z$ is the axial coordinate (for details see \cite{husakou_optexp_2009}%
). This equation does not rely on the slowly-varying envelope
approximation, includes dispersion to all orders, and can be used for the
description of extremely broad spectra. Since the transfer to higher-order
transfer modes is small, we consider only the fundamental linearly-polarized
HE$_{11}$-like mode. The nonlinear refractive index of argon is $n_{2}=1\times10^{-19}$ cm$%
^{2}$/W/atm. Besides the spectral and temporal properties, the coherence of the SC is of
crucial importance for most applications. We study the first-order coherence
function $g(\lambda )$ \cite{dudley_revmodphys_2006} which directly
corresponds to the visibility measured in interference experiments. The
effect of quantum noise is included by adding to the input field the shot
quantum noise in the approach of the Wigner
quasi-probability representation \cite%
{drummond_joptsocam_2001}.

\begin{figure}
\includegraphics{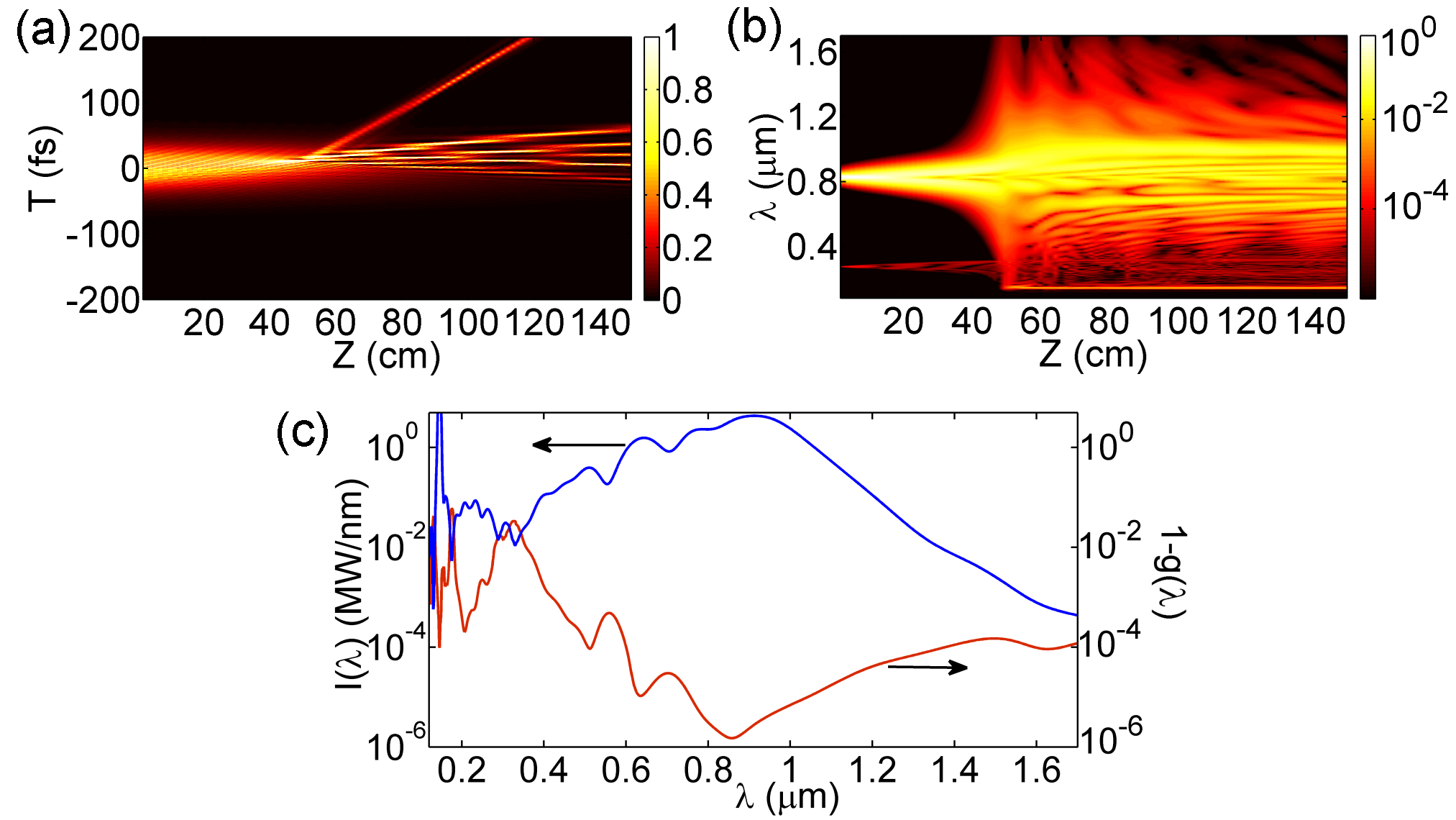}
\caption{High-power and high-coherence optical supercontinuum generation in
a kagome-lattice HC-PCF filled with argon at 0.25 atm. In (a) the evolution
of the temporal profile and in (b) the spectral evolution for an input 50-fs
pulse at 800 nm with a peak intensity of 176 TW/cm$^2$ are presented. In (c)
the output spectrum (blue upper curve) and the coherence function (red lower curve)
after 60 cm propagation are presented. The other parameters are as in Fig. \ref{fig:1_Kagome}.}
\label{fig:2_scg}
\end{figure}

The evolution of the temporal shape
presented in Fig. \ref{fig:2_scg}(a) can be divided into three stages: (i) the strong initial compression, (ii) 
formation of a high-order soliton with periodic modulations of the intensity,
and (iii) the final splitting into several fundamental solitons.
After 50 cm of propagation, besides the longitudinally 
modulated trace of a higher-order soliton, one can see a separate diverging trace.
 The presence of higher-order dispersion leads to a
transfer of energy from the high-order soliton to narrow-band resonant non-solitonic
radiation (NSR) in the normal GVD region, the position of this resonance is
determined by the phase-matching condition \cite{husakou_PRL_2001} which is 
the same as for the fundamental soliton. 
In contrast to solid-core PCFs, here
the absent Raman effect and low third-order dispersion coefficient result in
the stability of the high-order soliton over a considerable length.
This soliton-related dynamics explains the spectral evolution as presented in
Fig. \ref{fig:2_scg}(b). At the first stage of evolution the combined effect
of nonlinear phase modulation and anomalous dispersion 
results in a dramatic spectral broadening at around 50 cm (so-called soliton
compression) and the
formation of a high-order soliton. 
With further propagation, on the
long wavelength side a periodic modulation arises. The spectrum at the
optimum propagation length of 60 cm is shown in Fig. \ref{fig:2_scg}(c) and
cover more than two-octave broad range from 150 to 1300 nm. 
The spectral peak power density reaches high values up to the range of
MW/nm which is by five orders of magnitude higher than achieved previously
in solid-core PCFs. In the red lower curve of Fig. \ref%
{fig:2_scg}(c) we present the incoherence $1-g(\lambda )$ of the
supercontinuum, which is caused by the influence of quantum noise. It can be
seen that the coherence function $g(\lambda )$ is close to unity over the whole
spectral range, with the average spectrum-weighted value of $\overline{%
g(\lambda )}=1-10^{-4}$. The high coherence is explained by the relatively
low soliton number $N\sim10$; higher values of the soliton number
typically lead to a significant degradation of the coherence \cite%
{dudley_revmodphys_2006}. In a kagome-lattice HC-PCF the soliton number, and
therefore the output coherence, for the same input pulse parameters can be
controlled by the variation of the pressure.

\begin{figure}
\includegraphics{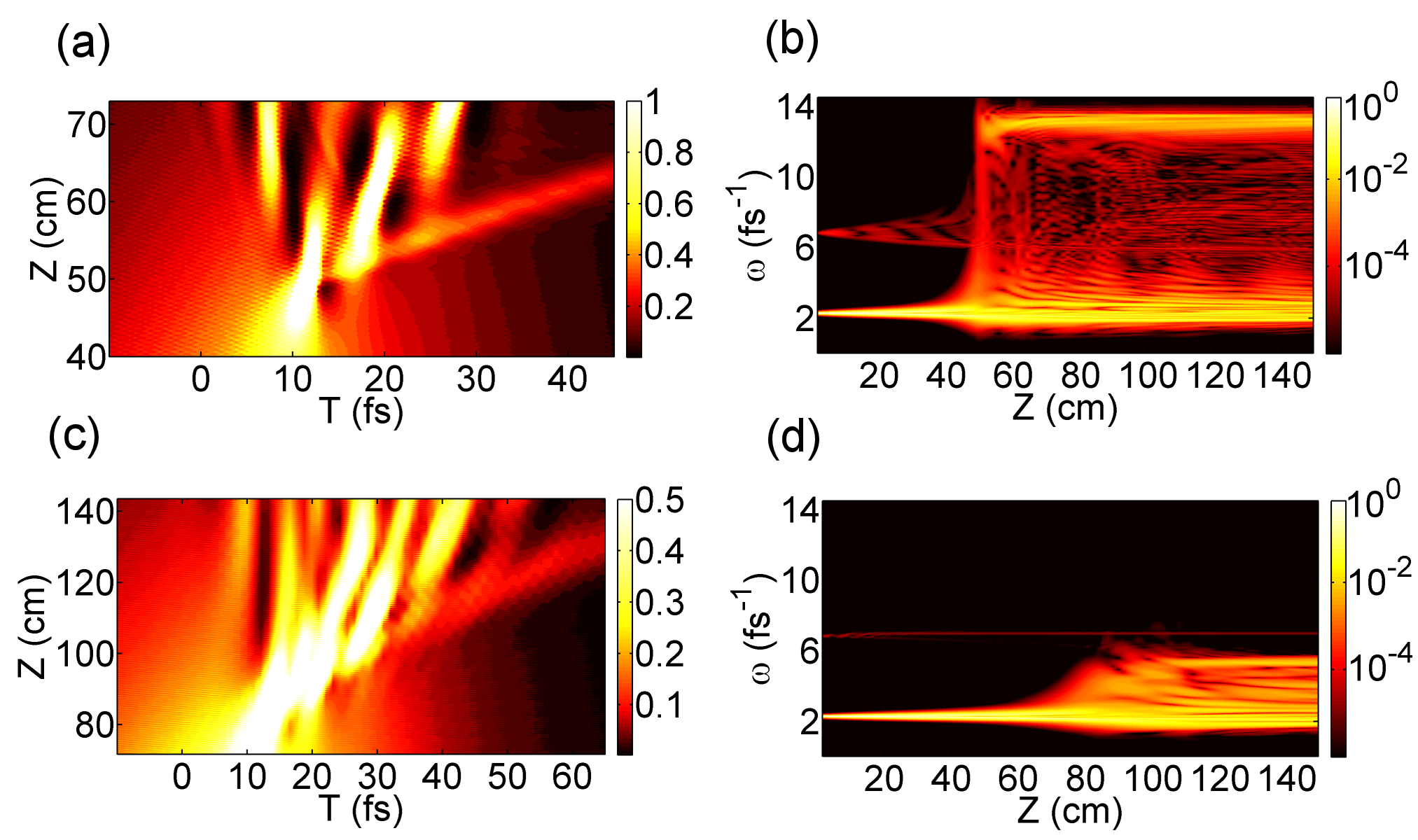}
\caption{Temporal and spectral evolution of generated radiations in the
kagome lattice HC-PCF filled with argon at different pressures. The input 50
fs pulse at 800 nm has the peak intensity of 176 TW/cm$^2$ for a pressure
of 0.25 atm (a),(b) and 14 TW/cm$^2$ for a pressure of 2 atm (c),(d).}
\label{fig:3_spec}
\end{figure}

An interesting physical phenomenon in Fig. \ref{fig:2_scg}(a) is connected
with the bright trace diverging at a large angle which appears in the
spectrum in Fig. \ref{fig:2_scg}(b) as the narrow-band bright trace at 150
nm. This isolated short-wavelength peak can be understood as arising by the
emission of NSR by the short-lived maximum-bandwidth stage of
the high-order soliton. At this stage, the overlap of the soliton spectrum with
the resonance frequency is maximal. 
This is more clearly seen
in the enlarged scale in Fig. \ref{fig:3_spec}(a) or in the evolution of
spectrum in Fig. \ref{fig:3_spec}(b) which is shown for clarity as a
function of frequency (the parameters are the same as in Fig. \ref{fig:2_scg}%
). In the first stage only the pump at 2.26 fs$^{-1}$ and its third harmonic
at 6.8 fs$^{-1}$ can be seen, but after 50 cm an intense 
band in the high-frequency region at 13 fs$^{-1}$ (corresponding to 150 nm)
is visible. 
The duration of the NSR peak is determined by a group velocity mismatch between the high-order soliton and the NSR and on the other hand, a propagation distance over which the soliton spectrum overlaps with the resonance frequency.
One can see in Fig. \ref{fig:2_scg}(a) and \ref{fig:3_spec}%
(a) that the NSR pulse is generated only as long as the soliton is strong enough and therefore broadband enough to provide seed components at the NSR frequency.
With further propagation, after roughly 3 cm the periodic modulation typical for a higher-order soliton leads to a reduction of the overlap and a disconnection of the soliton spectrum with the resonance frequency, and the generation of NSR stops, resulting in a NSR pulse duration below 10 fs. 
Careful examination of Fig. \ref{fig:3_spec} reveals
that the broadband soliton stage exists over roughly 3 cm, which in
connection with the group velocity mismatch of 6$\times 10^{-5}c$ leads to a 
NSR pulse duration of 6 fs, in correspondence
with the numerical observations. The above origin of the UV/VUV spectral
components can be checked by the study of the phase-matching condition. In
the solid (red) line in Fig. \ref{fig:1_Kagome}(d), the phase-matched
wavelength for the emission of NSR in dependence on the
argon pressure is presented. As can be seen, assuming a soliton 
 at 800 nm, the position of the resonance is at 150
nm for a gas pressure of 0.25 atm. The position of the phase-matched
wavelengths can be continuously tuned to longer wavelengths with increasing
pressure. To examine this prediction further, in Fig. \ref{fig:3_spec}(c),(d)
the spectral and temporal evolution of a 50-fs pulse at 800 nm for a
pressure at 2 atm is presented. The peak intensity here is only 14 TW/cm$^{2}$ 
and the bright high-frequency narrow
band at 5 fs$^{-1}$ (or 350 nm) is emitted at the wavelength corresponding
the phase-matching condition for 2 atm as given in Fig. \ref{fig:1_Kagome}%
(d). Note that the described emission of an ultrashort VUV pulse can not be
achieved in solid-core PCFs, where the extension of spectra below 350 nm is
impeded by loss in the VUV and much stronger dispersion which modifies the
fission dynamics and the energy transfer to the NSR.

\begin{figure}
\includegraphics{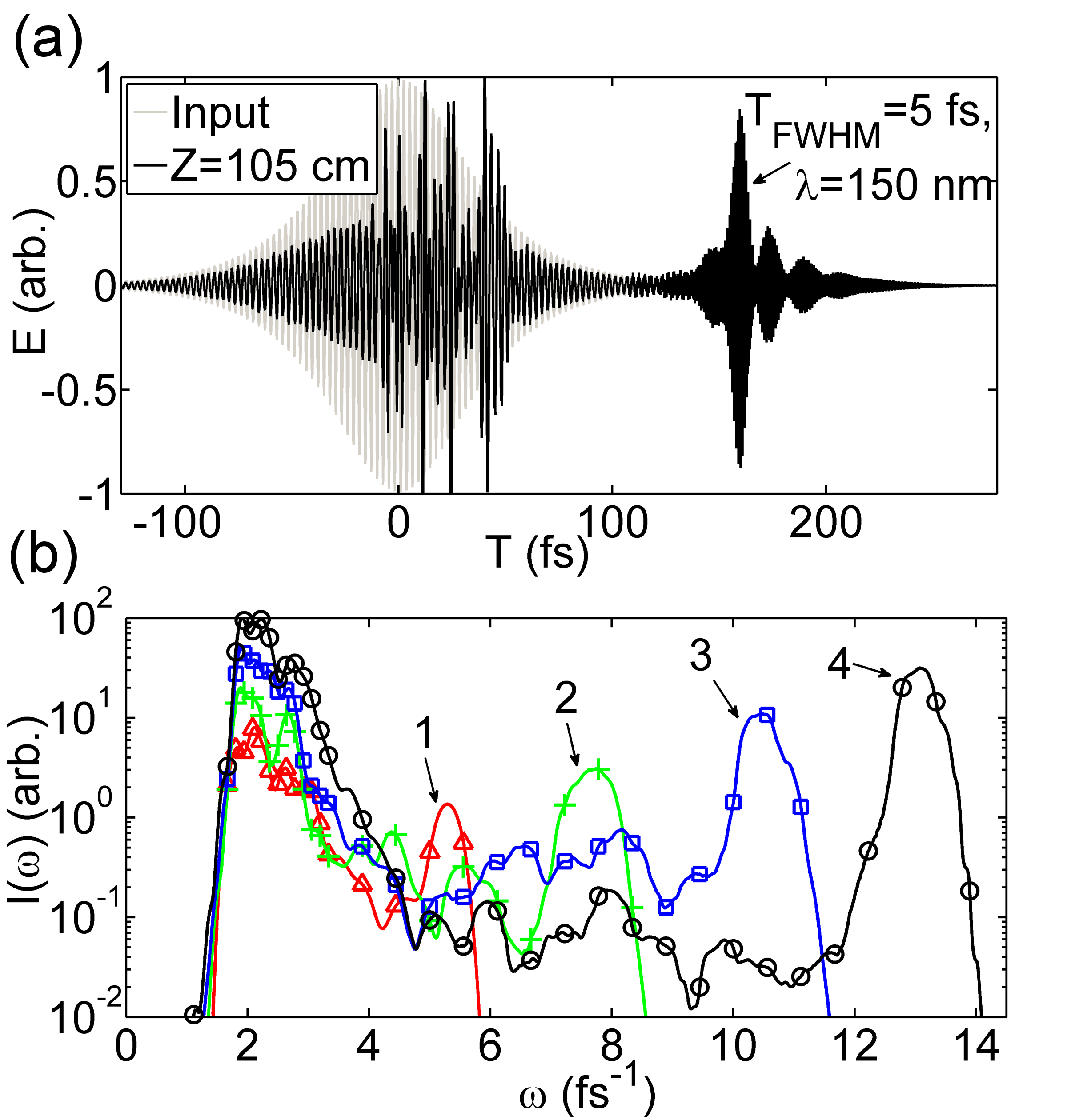}
\caption{Pulse shape (a) and tunable UV/VUV spectra for different pressures
(b). In (a) the electric field strengths for a pressure of 0.25 atm are presented. Here, the
gray solid represents the 50-fs input pulses at 800 nm and the black one the
electric field strength after propagation of 105 cm. In (b) the spectra after
propagation of 120 cm of the input 50-fs pulse at 800 nm at pressure of 2 atm and intensity of 14 TW/cm$^{2}$ (red curve 1), 1 atm and 35 TW/cm$^{2}$ (green curve 2), 0.5 atm and 88 TW/cm$^{2}$ (blue curve 3) and 0.25 atm and 176 TW/cm$^{2}$ (black curve 4) are presented.}
\label{fig:4_vuv}
\end{figure}

To study the emission of the bright UV/VUV component further, in 
Fig. \ref{fig:4_vuv}(a) the temporal shape of the electric field at a pressure of
0.25 atm after 105 cm propagation is presented. At this distance the bright
UV/VUV component with central wavelength at 150 nm is clearly separated from
the rest of the pulse, its duration (FWHM) is
only 5 fs and it exhibit about the same maximum intensity as the input
pulse. The energy of this ultrashort VUV pulse is about 25 $\mu$J or 20\% of the input pulse energy. The central frequency of the UV/VUV pulses can be simply tuned by the change of the pressure. As can be seen in
Fig. \ref{fig:4_vuv}(b) with decreasing pressure the high-frequency spectral
peak is tuned to shorter wavelengths; for 2 atm the UV/VUV component is at
350 nm, for 1 atm at 250 nm, for 0.5 atm at 180 nm and for 0.25 atm at 150 nm. For still lower
pressures the wavelength can be reduced up to 115 nm. Note that in the different curves of Fig. \ref%
{fig:4_vuv}(b) different input intensities has been
chosen, but the fraction of energy in the UV/VUV part is always about
20\% while the pulse durations vary from 5 fs for 0.25 atm to 20 fs for 2 atm,
due to different group velocity mismatch at different UV/VUV
frequencies.

In conclusion, we propose to use kagome-lattice HC-PCFs filled with a noble gas 
for the generation of high-power supecontinua and sub-10-fs pulses tunable by the
pressure over a broad UV-VUV range. These predictions
should be interesting for many  applications and can bring qualitative
advances in several fields. To remark a few: high power
SC sources can replace several lasers at different frequencies (including
frequencies where lasers do not exist), it can increase the detection sensitivity in nonlinear spectroscopic
methods such as CARS spectroscopy and bio-medical microscopy and
spectroscopy, and it can lead to advances in high-resolution frequency comb
spectroscopy. On the other hand, the prediction of a novel method for tunable
sub-10-fs VUV pulses with energy in the $\mu $J range could improve
the pulse parameters accessible ultrafast time-resolved
measurements in physics, chemistry, biology and other
fields.

We acknowledge financial support from the German Academic Exchange Service (DAAD) and the German Research Foundation (DFG).

\end{document}